\documentclass[%
 aip,
 amsmath,amssymb,
preprint,%
]{revtex4-1}

\draft 

\usepackage{graphicx}
\usepackage{dcolumn}
\usepackage{bm}

\usepackage[utf8]{inputenc}
\usepackage[T1]{fontenc}
\usepackage{mathptmx}

\begin{document}
\preprint{}

\title{In-situ grown single crystal aluminum as a non-alloyed ohmic contact to n-ZnSe by molecular beam epitaxy (MBE)}

\author{Zongjian Fan}
\email[]{zjfan@ucdavis.edu}
\affiliation{ 
Department of Electrical and Computer Engineering, University of California, Davis, Davis, CA 95616, USA
}

\author{Ryan Bunk}
\affiliation{ 
Department of Electrical and Computer Engineering, University of California, Davis, Davis, CA 95616, USA
}

\author{Guangying Wang}
\affiliation{ 
Department of Electrical and Computer Engineering, University of California, Davis, Davis, CA 95616, USA
}

\author{Jerry M. Woodall}
\affiliation{ 
Department of Electrical and Computer Engineering, University of California, Davis, Davis, CA 95616, USA
}

\date{\today}
\begin{abstract}
Novel ohmic contacts to n-ZnSe are demonstrated using single crystal Al films deposited on epitaxially grown ZnSe (100) by molecular beam epitaxy (MBE). Electron Backscatter Diffraction (EBSD) confirmed the single crystalline structure of the Al films. The (110)-oriented Al layer was rotated 45$^\circ$ relative to substrate to match the ZnSe (100) lattice constant. The as-grown Al-ZnSe contact exhibited nearly ideal ohmic electrical characteristics over a large doping range of n-ZnSe without any additional treatment. The contact resistances are in a range of 10$^{-3}$ $\Omega$-cm$^{2}$ for even lightly doped ZnSe ($\sim$10$^{17}$ cm$^{-3}$). Leaky Schottky behavior in lightly doped ZnSe samples suggested Al-ZnSe formed a low barrier height, Schottky limit contact. In-situ grown Al could act as a simple metal contact to n-ZnSe regardless of carrier concentration with lower resistance compared to other reported contacts in literatures. The reported novel metallization method could greatly simplify the ZnSe-based device fabrication complexity as well as lower the cost.\\
\end{abstract}
\maketitle

\section{Introduction}
ZnSe is a wide bandgap II-VI semiconductor extensively studied for blue-green laser diodes and light emitting diodes (LED)\cite{al1, al2, al3, al4}. More importantly, it is nearly lattice matched to GaAs, which is  widely used for many commercial photonic and electronic devices. There has been past interest in ZnSe-GaAs heterojunction and quantum well systems for devices such as full-colored LED pixelated displays, high power heterojunction bipolar transistor (HBT) and field-effect transistor (FET)\cite{al5, al6, al7, al8, al9, al10, arxiv}. More recently, ZnSe is considered as candidate material for quantum computation and communication as it can be made nuclear-spin free to obtain long electron spin coherence times\cite{al20}. For all these applications, low-resistance ohmic contacts are  essential for fabrication of high-performance devices.\\
However, ohmic contacts have been a challenge for many wide-gap II-VI compounds due to their poor thermal stability, including ZnSe\cite{al14}. Complicated graded structures like ZnS$_x$Se$_{1-x}$ or ZnSe$_x$Te$_{1-x}$ were employed  to facilitate ohmic contact formation\cite{al14, al11}. Previous studies have found many metal combinations like Ti/Pt/Au, Mg/Cu, In/Au could form ohmic contacts to n-ZnSe\cite{al11, al12}. However, most of them require a high doping level of ZnSe and additional treatments, such as plasma treatment or annealing, and suffer from high contact resistance\cite{al11, al12}. Indium (In) has also been used as an ohmic contact, but this contact has high resistivity and low reliability due to In’s low melting point and poor wetting\cite{al13}. Very recently, researchers also realized local ohmic contacts to a buried n-ZnSe layer through selective regrowth of heavily doped ZnSe layer, then deposited low work function metals such as aluminum (Al), magnesium (Mg) and titanium (Ti)\cite{al20}. \\
Single-crystal Al has been grown in-situ on GaAs (100) by Molecular Beam Epitaxy (MBE) at room temperature, and was found to form a Schottky rectifying contact\cite{al15}. Furthermore, researchers have shown that MBE-grown Al can form a leaky Schottky-based ohmic contact as well as a self-aligned mask for selective etching in GaAs solar cell fabrication\cite{al16}. Al was also used as a n-type dopant for ZnSe\cite{al17}, and could be embedded in many MBE systems for ZnSe growth. Moreover, it has close work function (4.06 – 4.26 eV) compared to the electron affinity of ZnSe (4.09 eV)\cite{al18, al19}. Therefore, MBE grown Al appears to be a good candidate for an ohmic contact to n-ZnSe.\\
In this work, we demonstrated that single crystal Al could be grown by MBE on ZnSe (100) surface at room temperature. In-situ MBE grown Al can serve as a uniform, low-resistance ohmic contact to n-ZnSe epi-layers without any additional treatment such as annealing or plasma. In addition, the Al to ZnSe contact exhibited low resistance over a large doping range of ZnSe and up to a high current density, even for the ZnSe samples which had been intentionally exposed to air. Compared to other reported ohmic contacts to n-ZnSe, Al has the lowest contact resistance and much simpler fabrication procedure, and could be applied to ZnSe layers with various carrier concentrations. This novel contact method is a significant improvement for the application of ZnSe in many electronic and photonic devices, and it has potential to be generalized to other II-VI material systems.\\

\begin{figure*}
\includegraphics[width=0.9\textwidth]{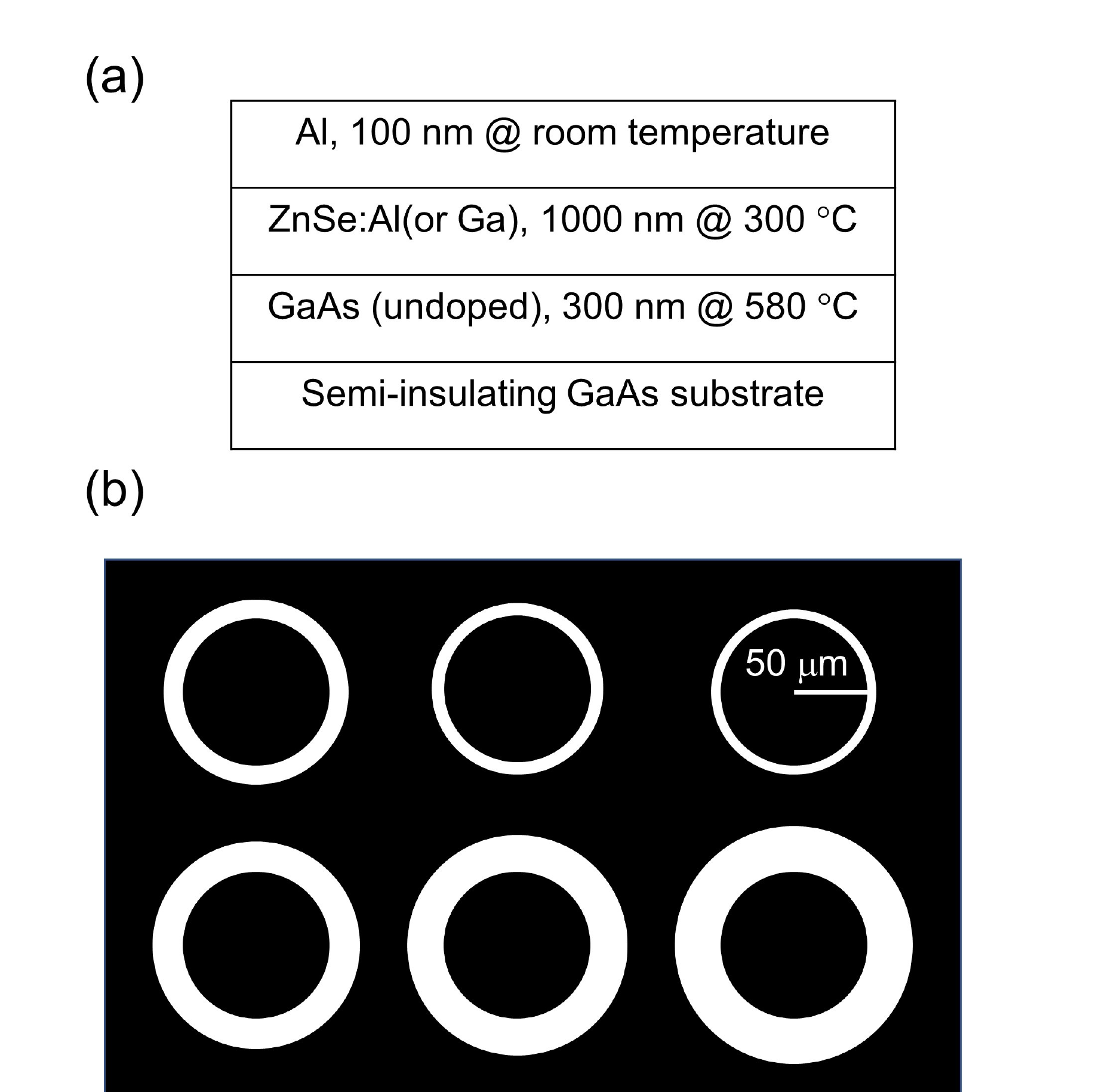}
\caption{\label{fig:epsart} (a) Sample stack structures for this study. (b) Schematic of the CTLM pattern. The inner circles have a radius of 50 $\mu$m, and the spacings between inner and outer circles are 5 $\mu$m, 10 $\mu$m, 15 $\mu$m, 20 $\mu$m, 25 $\mu$m, 30 $\mu$m, respectively.}
\end{figure*}

\section{Experimental}
All samples were grown in a Varian Gen II MBE system on semi-insulating GaAs (100) substrates. Figure 1a shows the typical sample stack structure. Prior to growth, the substrates were first baked in the load-lock chamber at a temperature of 300$^\circ$C for three hours to desorb water vapor. The GaAs substrates were then loaded in the growth chamber, and heated to $\sim$610$^\circ$C  under As overpressure to remove the native GaAs oxide layer. The thermal treatment process was continued until a dot diffraction pattern was confirmed by Reflection High Energy Electron Diffraction (RHEED). In order to provide a better-quality growth surface for ZnSe growth as well to electrically isolate the ZnSe layer, a 300 nm thick undoped GaAs buffer layer was grown on oxide removed GaAs substrates at a temperature of 580$^\circ$C, and with a V/III flux ratio of $\sim$15. Then a $\sim$1000 nm thick Al- or Ga-doped ZnSe was deposited in the same MBE growth chamber at 300$^\circ$C. Carrier concentrations in ZnSe epi-layers were measured via Differential Capacitance-voltage (CV) measurements. An effusive cell with compound ZnSe source material was used to provide the beam flux for ZnSe growth.\\ 
The details of samples fabricated for this study could be found in Table I. The 100 nm Al layers on differently doped ZnSe (sample A, B, C) were grown on room temperature (unheated substrate) in the same chamber after ZnSe growth, with a growth rate of $\sim$1 \AA/s. For comparison, one ZnSe sample was intentionally exposed to air for a day, and then reloaded into MBE system to grow Al under identical growth temperature and rate (Sample D). A circular transmission line (CTLM) pattern shown in Figure 1b was applied to all samples. HF:H$_2$O = 1:3 etchant was used to develop the patterns\cite{al15}. In addition, Ti/Au (20 nm/200 nm) layer grown by a e-beam evaporator with same CTLM patterns on ZnSe (sample E) was also fabricated for comparison to the in-situ Al contacts. \\
All I-V characteristics were measured by contacting the device with a probe station connected to a Keysight B1500A Semiconductor Parameter Analyzer. The Electron Backscatter Diffraction (EBSD) map was measured by a FEI Scios Dual Beam FIB/SEM system equipped with an Oxford EBSD detector. The specific contact resistances in Table I were calculated by methods described in a previous study\cite{ctlm}. The model and solutions with non-zero metal sheet resistance was used to avoid underestimation of contact resistance\cite{ctlm}.\\

\begin{table*}
\caption{\label{tab:table1}Properties of the various metal contacts on n-type ZnSe in this work (sample A-E) and literatures.}
\begin{ruledtabular}
\begin{tabular}{ccc}
Carrier concentration (cm$^{-3}$) & Contact metal \& conditions & Specific contact resistance ($\Omega$-cm$^{2}$)\\
\hline
5$\times$10$^{18}$	&	Al as grown (sample A)	&	2.433$\times$10$^{-3}$\\
9$\times$10$^{17}$	&	Al as grown (sample B)	&	6.911$\times$10$^{-3}$\\
2$\times$10$^{17}$	&	Al as grown (sample C)	&	1.05$\times$10$^{-2}$\\
5$\times$10$^{18}$	&	Al reload (sample D)	&		7.033$\times$10$^{-3}$\\
5$\times$10$^{18}$	&	Ti/Au as grown (sample E)	&	Not ohmic\\
1.15$\times$10$^{19}$	&	Mg/Au as grown\cite{al11}	&	9.95$\times$10$^{-2}$\\
1.15$\times$10$^{19}$	&	In/Au as grown\cite{al11}				&	1.04$\times$10$^{-2}$\\
4.5$\times$10$^{18}$	&	In/Au + 250$^\circ$C anneal\cite{al11}	&	 1.18\\
5$\times$10$^{16}$	&	Ti/Pt/Au as grown\cite{al12}				&	8.8$\times$10$^{-2}$\\
4$\times$10$^{18}$	&	Ti/Pt/Au as grown\cite{al12}				&	6.2$\times$10$^{-2}$\\
2$\times$10$^{19}$	&	Ti/Pt/Au as grown\cite{al12}				&	3.4$\times$10$^{-4}$\\
2$\times$10$^{18}$	&	In + 200$^\circ$C anneal\cite{al13}	&	5$\times$10$^{-2}$\\
\end{tabular}
\end{ruledtabular}
\end{table*}

\section{Results and Discussion}
Previous studies have shown Al could be grown epitaxially on GaAs at both elevated and room temperatures\cite{al15, al21, al22, al23}. In this study, Al films were grown at room temperature to form the contact to n-ZnSe. As shown in Figure 2a, the ZnSe as grown (100) surface showed streaky RHEED pattern before Al growth, indicating an atomically flat surface. After initializing the Al growth, the RHEED pattern changed to  spotty within first 3 minutes (corresponding to $\sim$20 nm Al), indicating a 3D growth mode. During the very first few \AA, some additional diffraction spots were observed, which was also mentioned in previous study on Al on GaAs growth\cite{al15}. That could be caused by the change of Al lattice at the initial growth. The RHEED pattern turned streaky gradually and ended up with (3$\times$1) surface reconstruction after a while (Figure 2c and 2d), which suggested the Al growth surface was highly perfect. Another point to be noted from the RHEED patterns is that the spacings between diffracted beams for Al and ZnSe were nearly the same (indicated by the arrows in Figure 2 with same separation distance). Previous studies also observed same diffracted beam spacings for Al on GaAs growth, and they proposed that a rotation of the face-center cubic (FCC) Al unit cell of 45$^\circ$ about the c axis will closely match the underlying GaAs lattice\cite{al15, al21, al22}. Since ZnSe has very similar lattice constant with GaAs, we believe the same phenomenon happened during Al on ZnSe growth. To confirm that, EBSD was performed on the surface of Al film, and the orientation was confirmed to be (110)//(100) (Al/ZnSe) (Figure 2e).. The orientation distribution map  showed no divergence over the whole surface for all three directions, which further confirmed the single crystalline of Al film. Since the lattice plane spacing between d$_{[220]}$ Al and d$_{[400]}$ ZnSe only has a mismatch of 0.56\%, it reasonable that the Al lattice could be rotated to match the ZnSe (100) lattice constant. The transient RHEED pattern mentioned above during initial growth was likely due to the lattice rotation. Identically (110)-oriented Al  film was also observed during the Al on As-rich GaAs (100) growth at room temperature\cite{al21, al22}.\\

\begin{figure*}
\includegraphics[width=0.9\textwidth]{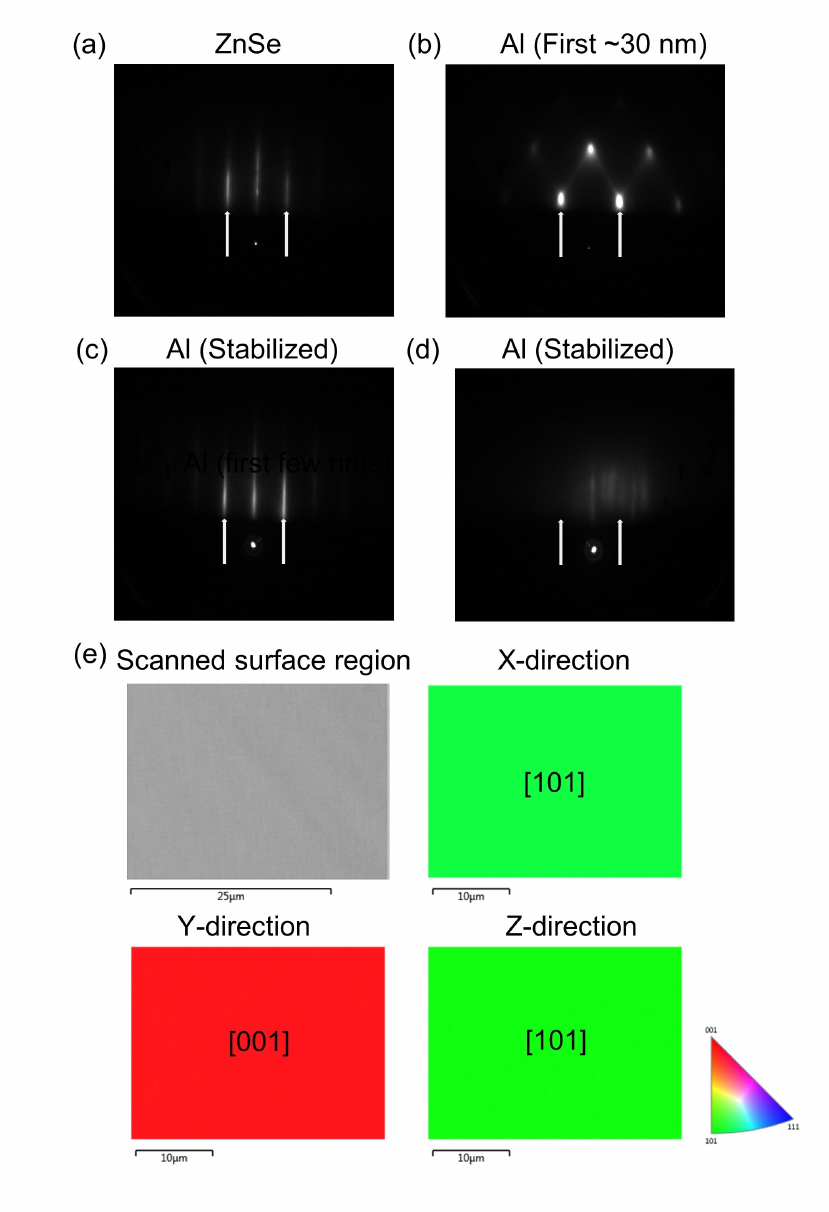}
\caption{\label{fig:epsart}RHEED patterns of (a) MBE ZnSe (100) as-grown surface, (b) after depositing 30 nm of Al on the MBE ZnSe. And (c), (d) shows RHEED patterns of Al when stabilized from different coordinates. The arrows indicate the positions of the diffraction from the bulk. (e) shows the orientation distribution map of Al surface by EBSD, z-axis is perpendicular to the growth surface. Color code: crystal axis along the axis direction.}
\end{figure*}

The Al ohmic contacts used in this study were single-crystal Al grown on n-type ZnSe MBE layers, doped with Ga or Al with different carrier concentrations (sample A-C). Using MBE for in-situ contact deposition, one could obtain an impurity-free metal-semiconductor interface, which is important for both device performance and scientific research\cite{al15}. For comparison, one ZnSe sample (sample D) was intentionally exposed to air for a day and then reloaded into MBE system for Al deposition. Ti/Au on n-ZnSe sample (sample E) was also fabricated as a traditional method to form ohmic contact for comparison study, since Ti/Pt/Au showed the lowest contact resistance among available literatures\cite{al12}. The I-V characteristics of as-grown Al, reload-and-grown Al, and Ti/Au on N-ZnSe layers were shown in Figure 3a-3d. It could be seen that all Al on ZnSe samples showed nearly ideal linear characteristics over a large current density range (up to 390 A/cm$^{2}$). Especially, all Al on ZnSe samples exhibited great linearity in low operation voltage region. However, the Ti/Au on ZnSe sample showed rectifying (blocking) behavior, suggesting the existence of Schottky barrier blocking electron flows. Previous study showed as-deposited Ti/Pt/Au (10 nm/100 nm/200 nm) could form ohmic contact to n-ZnSe with an electron concentration greater than 10$^{19}$ cm$^{-3}$, and researchers attributed this to the interface interdiffusion between Ti and ZnSe, as well as the low work function of Ti\cite{al12}. Since Ti was identified as the key role for ohmic contact, in this study, Ti/Au (20 nm/200 nm) was deposited on n-ZnSe with a carrier concentration of 5$\times$10$^{18}$ cm$^{-3}$ (sample E), but it failed to form an ohmic contact. This could be due to the lower doping level used in this study compared to previous one, or different content of interdiffusion due to fabrication procedures, or that a mechanism other than Ti plays the key role in the Ti/Pt/Au ohmic contact formation. Table I summarizes the properties of ohmic contacts in this study and previous literatures. It should be noted that Al contact has order of magnitude lower contact resistance compared to other known metal contacts with similar carrier concentrations in ZnSe. Interestingly, the contact resistance did not degrade for lightly doped ZnSe. In addition, there is no need for additional treatments, such as thermal annealing to form the Al ohmic contact. The in-situ metallization method using a low-cost single metal Al could greatly simplify the device fabrication procedure and save product costs. \\

\begin{figure*}
\includegraphics[width=0.9\textwidth]{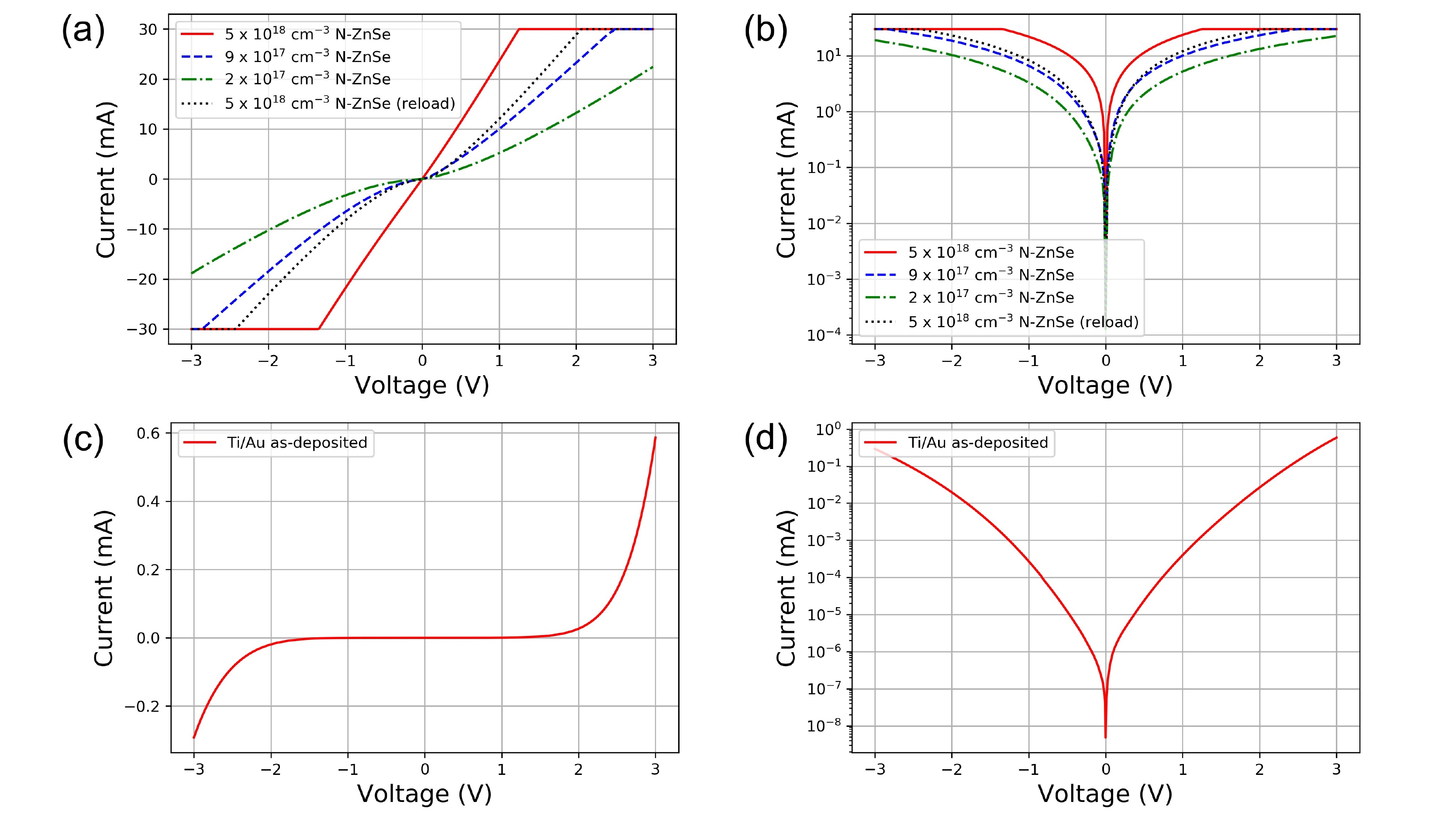}
\caption{\label{fig:epsart}I-V characteristics of (a) Al (sample A - D) and (c) Ti/Au (sample E) contact to ZnSe. The current limit set in (a) was 30 mA. (b) and (d) shows the corresponding I-V in semilog scale. All data was measured from the circle with largest spacing on the CTLM pattern. }
\end{figure*}

Theoretically, abrupt metal-semiconductor contacts can be described by the Schottky barrier model\cite{al24}. In practice, several non-ideal mechanisms including surface state, fermi-level pinning, and interface reactions make the measured barrier heights strongly diverge from the theoretical values\cite{al25, al26}. Previous studies obtained a barrier height of 0.66 eV for Al on n-type As-rich GaAs surface by MBE, and 0.72 eV for Ga-rich GaAs\cite{al15}. It is well-known the fermi level of n-type GaAs (100) as-grown surface will be strongly pinned at mid-gap due to formation of surface acceptor states during MBE growth\cite{al27}. The extracted barrier height values in previous study clearly demonstrate that the measured barrier height is intrinsically determined by fermi level pinning position\cite{al25, al27}, i.e. different surface stoichiometries of GaAs, which follows the Bardeen model\cite{al28}. In this study, the work functions of Al are comparable to the ZnSe electron affinity\cite{al18, al19}. Therefore, it would form a low-barrier Schottky contact if it follows the Schottky model. The slightly nonlinear I-V characteristics of Al on ZnSe layers with lower carrier concentrations or exposed to air suggested the existence thermionic emission and/or quantum tunneling mechanism, indicating it behaves as a leaky Schottky contact with a very low barrier height. Besides, the contact resistances increased slightly with decrease of carrier concentrations, suggesting a higher barrier height for lower doped n-ZnSe. The result agrees with that Al-ZnSe is a Schottky limit contact. However, another metal combination (Ti/Au) would result in a rectifying contact to the same ZnSe surface (exposed to air). Since Ti has similar work (4.33 eV) function compared to Al, it seems contradict with the Schottky limit assumption. But a more recent work also showed that ex-situ deposited Ti on n-ZnSe with 2$\times$10$^{19}$ cm$^{-3}$ carrier concentration exhibited rectifying behavior, and researchers attributed this to the ZnSe surface oxide after being exposed to air\cite{al20}. Therefore, the in-situ deposited Al on ZnSe could still follow the Schottky model. Furthermore, the low barrier height Schottky contact formed between Al and lightly doped ZnSe samples suggested there was no significant mid-gap fermi level pinning at n-ZnSe surface (as grown and air exposed), unlike GaAs mentioned above\cite{al15}. More interestingly, previous studies observed ex-situ deposited Al on heavily doped n-ZnSe (2$\times$10$^{19}$ cm$^{-3}$) by a metal evaporator showed rectifying behavior, like Ti\cite{al20}, which appeared that the fermi level of air-exposed ZnSe surface will be re-pinned at mid-gap. However, in this study, MBE grown Al on air-exposed ZnSe with lower carrier concentration still exhibited a very low barrier height leaky Schottky characteristic. It’s reasonable to expect that polycrystalline Al by evaporator could have different interface behavior compared to epitaxially grown single crystalline Al, or the epitaxial growth procedure could alter the surface properties of ZnSe. Further experiments will be required to illuminate the difference between the epitaxial and polycrystalline interface.\\

\section{Conlusions}
In summary, we demonstrated in-situ MBE growth of single-crystal Al on n-ZnSe as a non-alloyed, low-resistance ohmic contact: (1) 2D growth mode with surface reconstruction was observed by RHEED during Al growth, and EBSD confirmed that Al lattice was rotated by 45$^\circ$ relative to substrate to (110)-oriented in order to match the ZnSe (100) lattice constant; (2) I-V characteristics of Al on heavily n-ZnSe layers showed ideal ohmic characteristics, and similar contact resistances from n-ZnSe with different carrier concentrations were observed; (3) Leaky Schottky I-V characteristics for Al on lightly doped ZnSe samples suggests the Al-ZnSe could form a low barrier height Schottky contact, and there is no mid-gap fermi level pinning at n-ZnSe surface (both as-grown and air exposed), which is different from n-GaAs. In-situ grown Al on n-ZnSe could be a Schottky limit contact; (4) Comparison with recent work\cite{al20} showed polycrystalline Al by evaporator and epitaxially single crystalline Al have  different interface behaviors. More investigation will be performed to further explain the nature of Al/ZnSe interface.\\
MBE grown Al was found to  form ohmic contact to n-ZnSe without any additional treatment like plasma or annealing used for traditional ohmic contacts. It has lower contact resistance (up to 2.433$\times$10$^{-3}$ $\Omega$-cm$^2$) compared to all current available metal contacts to n-ZnSe with similar carrier concentration. Moreover, the low contact resistance doesn’t rely much on the carrier concentration of bulk ZnSe, which could enable application to circumstance that requires lightly doped material. This method could greatly reduce the fabrication complexity and cost for ZnSe-based device applications. Furthermore, it could be generalized to solve the common ohmic contact difficulty for other wide-gap II-VI materials with similar electron affinity.\\

\textbf{AVAILABILITY OF DATA}\\
The raw data that support the findings of this study are available from the corresponding author upon reasonable request.\\

\bibliography{aipsamp}
\end{document}